\documentclass{aa}  
\usepackage{graphicx}
\usepackage[varg]{txfonts}
\usepackage{color}
\usepackage{natbib,twoopt}
\usepackage[colorlinks=true,
            linkcolor=blue,
            urlcolor=blue,
            citecolor=blue]{hyperref}
\bibpunct{(}{)}{;}{a}{}{,} 

\begin{document}

   \title{Using near infra-red spectroscopy for characterization of \\transiting exoplanets}
   \subtitle{}

   \author{E. Aronson
          \and P. Waldén
          \and N. Piskunov
          }
   \institute{Observational Astronomy, Department of Physics and Astronomy, Uppsala University, Box 516, SE-751 20 Uppsala, Sweden.\\
              \email{erik.aronson@physics.uu.se}
              }
   \date{Received April 24, 2014; accepted March 30, 2015}
%
 
  \abstract
   {We propose a method for observing transiting exoplanets with near-infrared high-resolution spectrometers.}
   {We aim to create a robust data analysis method for recovering atmospheric transmission spectra from transiting exoplanets over a wide wavelength range in the near infrared.}
   {By using an inverse method approach, combined with stellar models and telluric transmission spectra, the method recovers the transiting exoplanet's atmospheric transmittance at high precision over a wide wavelength range. We describe our method and have tested it by simulating observations.}
   {This method is capable of recovering transmission spectra of high enough accuracy to identify absorption features from molecules such as O$_2$, CH$_4$, CO$_2$, and H$_2$O. This accuracy is achievable for Jupiter-size exoplanets at S/N that can be reached for 8m class telescopes using high-resolution spectrometers (R > 20~000) during a single transit, and for Earth-size planets and super-Earths transiting late K or M dwarf stars at S/N reachable during observations of less than 10 transits. We also analyse potential error sources to show the robustness of the method.}
   {Detection and characterization of atmospheres of both Jupiter-size planets and smaller rocky planets looks promising using this set-up.}


   \maketitle
%

\section{Introduction} \label{Introduction}
The next big steps in exoplanetary science is to investigate the physical conditions on worlds outside our solar system. One of the most promising methods for this is so-called transit spectroscopy, which makes it possible to determine physical parameters of the exoplanetary atmosphere, most noticeably the chemical composition, but possibly also pressure and temperature profiles, rotation period, global weather patterns, cloud coverage, etc.

From radial velocity measurements and transit light curves, orbital parameters, radius, mass, and mean density of an exoplanet can be calculated \citep{Winn2010arXiv1001}. With this, one can divide exoplanets into categories, such as gas giants, ice giants, and rocky planets. A large number of the discovered exoplanets are categorized as hot-Jupiters, which are gas giants orbiting their stars often at distances closer than Mercury to the Sun. The ubiquity of such objects represents a challenge to our planetary-system formation theories \citep{Boss1997Sci, Nagasawa2008ApJ, Coleman2014MNRAS}. Another problem is the discrepancy between mass and radius for many smaller exoplanets, which falls between the rocky and gas planets of the solar system. Several possible solutions have been proposed, such as ocean planets, rocky planets with large extended atmospheres, or metal-poor planetary cores (similar to that of the moon), but the true nature of these worlds remains unclear \citep{Leger2004Icar, Castan2011ApJ, Rogers2010ApJ}.

Monitoring stellar brightness over long periods of time in search of exoplanet transits (i.e. when the exoplanet moves in front of its host star from our points of view) has proven to be a very successful detection method. Since the first observed transit event \citep{Charbonneau2000}, more than one thousand (November 2014) transiting exoplanets have been discovered. The success of the transit method for exoplanet detection comes mainly from the ability to observe many potential systems simultaneously. 

The feasibility of exoatmosphere detection by transit observations has been demonstrated by several authors \citep{Brown2001ApJ, Hubbard2001ApJ, Seager2000ApJ}. During the transit, stellar light can pass directly though the semi-transparent part of the atmosphere. The low contrast ratio between an exoplanetary atmosphere and its host star makes detecting the atmosphere a challenging task. The projection of a planetary atmosphere for a solar-size star covers of the order of $10^{-3}$ of the stellar disk for Jupiter-size planets, and can be as low as $10^{-6}$ for super-Earths and Earth-size planets.

Many successful detections using transit spectroscopy have been made. The two most commonly used methods for obtaining transit spectra are (1) low-resolution spectroscopy or photometric measurements of the transit light curve in filters corresponding to typical broadband absorption features, which was first demonstrated by \citet{Jha2000ApJ}; and (2) high-resolution narrow-band spectroscopy targeting a specific absorption line (or band of lines), first demonstrated by \citet{Charbonneau2002}. The transit curves from the first method are used to calculate the radius of the planet as a function of wavelength. Differences in perceived radius of a planet in different filters are interpreted as the presence of a semi-transparent atmosphere. This method makes it possible to recreate the transmission spectrum over a wide wavelength range, albeit at low spectral resolution. Molecules responsible for the observed absorption features typically can not be identified directly due to low resolution. Instead one usually tries to find the best fitting model of the planetary atmosphere based on a library of such models. The second method  depends less on models. Here one targets an absorption line of a molecule assumed to be present in the atmosphere of a planet. By measuring the relative intensity dip at wavelengths corresponding to this line and the intensity at wavelengths close to this, the presence of the molecule can be detected. Both these methods are applicable to ground- and space-based observations, although space offers a number of obvious advantages.

One of the best studied hot-Jupiters is \object{HD 209485 b}, which is slightly larger than Jupiter ($R_p = 1.36 R_{Jup}$) and which orbits a bright (V mag = 7.7) solar-like star. HST STIS spectra of this exoplanet during transit has led to the detection of sodium \citep{Charbonneau2002}, atomic hydrogen \citep{Vidal-Madjar2003Natur, Ballester2007Natur}, oxygen, carbon \citep{Vidal-Madjar2004ApJ}, magnesium \citep{Vidal-Madjar2013A&A}, and molecular hydrogen \citep{Lecavelier2008}. Using a new observational strategy (ground-based high-resolution spectroscopy, CRIRES at VLT) and data analysis method, \citet{Snellen2010} detected CO absorption in the exoplanet's atmosphere. Another well-studied hot-Jupiter is \object{HD 189733 b}, for which methane and water \citep{Swain2008Natur} were detected using HST observations, and sodium was detected through high-resolution ground-based observation (HRS at HET) \citep{Redfield2008ApJ}. 

Smaller exoplanets require smaller host stars such that $R_{star}/R_{planet}$ favours detection of exoplanetary atmospheres. The atmospheric transmittance of the Neptune-size exoplanet \object{GJ 436 b} has been measured, revealing a mostly featureless spectrum in near infra-red (NIR) \citep{Pont2009MNRAS, Knutson2014Natur}.

The most extensively studied super-Earth is \object{GJ 1214 b} \citep{Bean2010, Bean2011ApJ, Carter2011ApJ, Croll2011, Crossfield2011ApJ, desert2011ApJ, Berta2012ApJ, Murgas2012A&A, deMooij2012AA, deMooij2013ApJ, Colon2013ApJ, Fraine2013ApJ, Narita2013ApJ, Teske2013MNRAS, Wilson2014MNRAS, Caceres2014A&A, Kreidberg2014Natur}. The small host star (an M dwarf with $\mathrm{R_{star} = 0.21 R_{\odot}}$) gives the planet a favourable contrast ratio. The combined wavelength coverage from these studies ranges from 0.4 to 5 $\mu m$.  Most studies have concluded that GJ 1214 b has a flat and featureless transmission spectrum, usually suggested to be caused either by a water-dominated atmosphere or by high altitude clouds or hazes.

These efforts have shown that with current methods and instruments, it is indeed possible to detect exoplanetary transmission spectra. The strong model dependence of the fitting procedure may, however, introduce uncertainties and degeneracies in the analysis of the atmosphere. The goal of this paper is to present an alternative observational strategy and data analysis that is capable of obtaining exoplanetary atmospheric transmittance over a wide wavelength range with sufficient resolution to identify the sources of spectral features directly from their shapes and placements. From such a spectrum, the chemical composition and physical conditions in the atmospheres could be deduced with less model dependence. Examples in this paper are in NIR due to our interest and involvement in the CRIRES+ upgrade (see Sec. \ref{Instruments}), but we foresee no reasons that this could not be applied to other instruments or wavelength domains. We also formulate the instrumental requirements needed to obtain suitable observations for this method. 


\section{Method} \label{Method} 
Here we describe the data analysis method for ground-based high-resolution transit spectroscopy.


\subsection{Synthetic observations} \label{Synthetic observations} 
To test the data analysis method, we need observations of transiting exoplanets. To have control over the data and to know the accuracy of the reconstructed exoplanet transmission spectrum, we create artificial observations with a hypothetical high-resolution spectrograph. This way we have full control over the signal-to-noise ratio (S/N), the ratio of planetary-to-stellar radius, atmospheric conditions, radial velocity, etc., and thus are able to reliably estimate the requirements for instruments and target selection. The simulated observations are based on several assumptions, as follows. 
\begin{enumerate}
\item A solid (or cloud-covered) part of the planet blocks stellar light completely in the observed wavelengths, while the atmosphere above absorbs light as a function of wavelength.
\item The exposure times are short enough to ignore the motion of the planet across the stellar disk during a single exposure.
\item The stellar flux spectrum and exoplanet transmission spectrum remain constant during the whole transit.
\item The noise of observations is dominated by photon statistics.
\end{enumerate}

\noindent We simulate several short exposures of the transiting system as the planet moves across the stellar disk. A simulated spectrum ($\tilde{S}_n$) of a star with a transiting exoplanet reaching our detector is computed as
\begin{align} \label{eqq1}
& \tilde{S}_n(\lambda, t) = \Bigg[ F(\lambda, v_{s}) - i_p(\lambda, \phi, v_{s}) \frac{R_p^2}{R_{\star}^2}  - i_a(\lambda, \phi, v_{s}) P(\lambda, v_{s+p}) \Bigg] \cdot T(\lambda, t)\\ \nonumber
& i_p(\lambda, \phi, v_{s}) = \int\limits_{A_p(\phi)}  I(\lambda, \mu, v_{s}) d\mu \\ \nonumber
& i_a(\lambda, \phi, v_{s}) = \int\limits_{A_a(\phi)}  I(\lambda, \mu, v_{s}) d\mu \\ \nonumber
& P(\lambda) = \frac{H(\lambda)^2+2R_pH(\lambda)}{R_{\star}^2} \\ \nonumber
\end{align}
The expression in square brackets replaces the contribution of the obscured area on the stellar surface with the stellar light that passes through the exoplanetary atmosphere. Here $F$ is the flux of the star, and $i_p$ and $i_a$ are the specific intensities at the position of the planet or its atmosphere. These are given by integrating the specific intensity ($I$, characterized by the normalized limb distance, $\mu$,) over the area of the opaque part of the planet ($A_p$) or the semi-transparent atmosphere ($A_a$). If the planet moves significantly over the stellar disk during an exposure, the integration can be used to represent this movement, integrating over the motion across the stellar disk. $R_p^2/R_\star^2$ is the fraction of the stellar disk that is obscured by the opaque part of the planet. 

Our goal is to recover the transmittance of the exoplanet atmosphere, represented by $P$, where $P$ is calculated using the wavelength dependent height of the opaque atmosphere $H$ and is rescaled in units of stellar disk area. This means that $P$ is the fraction of the stellar disk that is blocked by absorption in the atmosphere. Telluric transmittance is given by $T$, and because of temporal variations (represented by $t$), we allow $T$ to change with each exposure. Simulations of observed spectra are made for many short exposures, which are represented by subscript $n$. Each exposure is taken during a given point in the orbit of the exoplanet, represented by phase $\phi$, which in turn corresponds to different $\mu$ as the exoplanet moves across the stellar disk during the transit. The rotation of the Earth (of the order of 0.5 km/s), and the star-planet system's velocity relative to Earth will result in a small Doppler shift in the stellar spectrum relative to the instrument, represented by $v_s$. The exoplanet will also move in its orbit during the transit. The radial velocity of a planet in transit is generally small, but for planets orbiting very close to their stars, variations during the transit produces non-negligible red and blue shifts, which is represented by $v_p$. All velocities are given relative to the detector.

\begin{figure}
\centering
\includegraphics[width=\hsize]{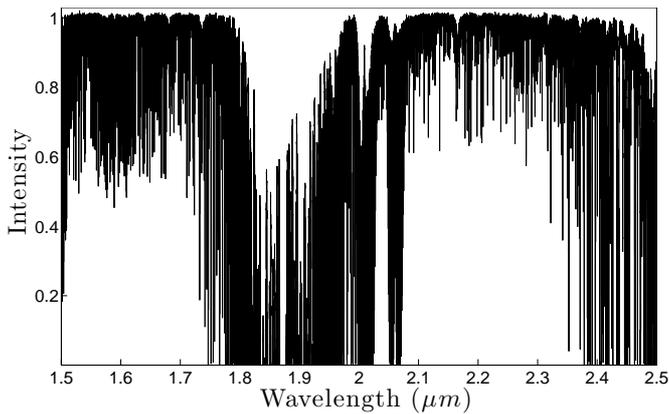}
\caption{Simulated observation of a hot-Jupiter transiting solar-like star, at S/N 175 and spectral resolution R  = 100~000. All visible lines have telluric or stellar origins, and the planet's contribution is not visible at this scaling since it is at the level of $10^{-3}$ (for the strongest lines) of stellar contribution.}
\label{example of observation}
\end{figure}

The resulting spectrum is convolved with an instrumental profile ($\Gamma$), which is approximated by a Gaussian with the FWHM given by the spectral resolution of the instrument (which we assume to be constant over the entire wavelength range). We then add white noise to each pixel. The realization of the noise ($N$) for a certain exposure is given by Poisson distribution, with the width of the distribution given by the inverse of the S/N selected for the data. We have ignored the effect of the blaze on the S/N within a spectral order. Finally, the spectrum is continuum normalized ($\eta$), and with this the full observation is created:
\begin{equation} \label{eqq2}
S_n(\lambda, t) = \big[\tilde{S}_n(\lambda, t) \textstyle \bigotimes \Gamma \cdot (1-N(\lambda , \phi)) \big]  \cdot \eta
.\end{equation}
To take the temporal changes in telluric transmittance into account accurately, as well as to use specific intensity corresponding to the different parts of the stellar disk that the planet covers during the transit, it is preferable to use many short exposures (0.5 - 2 min) of the transit rather than a single long integration. An example of a synthetic observation is given in Fig. \ref{example of observation}. 

\subsection{Data analysis method} \label{Data analysis method} 
Once simulated observations are produced, we may use them to reconstruct the transmission spectrum of the exoplanet ($P$). The straightforward approach would be to rewrite Eqs. \ref{eqq1} and \ref{eqq2} to express $P$ directly:
\begin{align} \label{eqq3}
 P(\lambda) \approx \frac{1}{n} \displaystyle \sum_{n} & \Bigg[\frac{-S_n(\lambda , t, -v_{s+p}) \cdot \nu(\phi)}{\big( T(\lambda, t, -v_{s+p}) \cdot i_a(\lambda, \phi, -v_{p}) \big) \bigotimes \Gamma} + \\ \nonumber
& \frac{ F(\lambda, -v_{p}) - R_p^2/R_{\star}^2 i_p(\lambda, \phi, -v_{p})}{i_a(\lambda, \phi, -v_{p}) } \textstyle \bigotimes \Gamma \Bigg].
\end{align}
Summation is carried out over many short exposures taken during the transit and then divided by the total number of exposures. Before adding spectra, all spectra are shifted to the reference frame of the exoplanet. Solving Eq. \ref{eqq3}, however, fails to recover the exoplanet spectrum since the resulting reconstructed planetary spectrum is swamped by noise amplified in wavelength regions where the denominator (usually telluric transmittance) is small. Rebinning the data to a lower resolution will not help because we will lose the information available between the telluric lines.

An alternative approach is to formulate and solve an inverse problem. Instead of using the expression for $P$ explicitly, we formulate an optimization problem, where we are searching for the function $P$ that minimizes the expression below:
\begin{align} \label{eqq4}
\Phi \equiv \sum_{n \lambda} & \Bigg\{
P(\lambda, v_0) \cdot 
\Big[ T(\lambda, t, -v_{s+p}) \cdot i_a(\lambda, \phi, -v_{p}) \Big] 
\textstyle \bigotimes \Gamma + \\ \nonumber
& S_n(\lambda, t, -v_{s+p}) \cdot \nu(t) - 
\Big[ F(\lambda, -v_{p})  \cdot T(\lambda, t, -v_{s+p}) + \\ \nonumber
& R_p^2/R_{\star}^2 i_p(\lambda, \phi, -v_{p}) \cdot T(\lambda, t, -v_{s+p}) \Big] \textstyle \bigotimes \Gamma \Bigg\}^2 = min.
\end{align}
This is essentially a rewritten version of Eqs. \ref{eqq1} and \ref{eqq2}. To be able to combine spectra, we need all spectra to be on the same wavelength grid. Observations ($S$), telluric transmittance ($T$), stellar flux ($F$), and specific intensity ($i$) are shifted and interpolated onto the rest frame of the exoplanet. These changes in the wavelength grid are represented by $v_s$ and $v_p$ for stellar radial velocity relative to observer ($v_s$) and exoplanet radial orbital velocity relative to the star ($v_p$). Exoplanet rest frame is used for establishing the wavelength vector, and we assume that $P$ is constant during the transit. We also assume that the observed spectra is continuum-normalized, but a perfect normalization will differ for the spectra in and outside the transit, accounting for the change in stellar flux. This problem should be addressed from two sides. On the data reduction side, we can use a robust and reproducible algorithm. We have developed such a tool and tested it with a number of fibre-fed and slit spectrographs, including ESO HARPS and UVES and the Keck HIRES. The main idea is to observe the solar spectrum (e.g. reflected from an asteroid) before and after the science exposures keeping instrument configuration. Sensitivity curves derived from the solar spectrum are then used to normalize the science exposures. In the absence of large seeing variations, this techniques provides precise continuum normalization at the level of $10^{-3}$. In addition, we introduce the brightness correction parameter $\nu$ on the modelling
side. This parameter primarily depends on stellar limb darkening and exoplanet radius. The optimal value of $\nu$ is found by calculating the residuals ($r$ in Eq. \ref{renorm}) between observations multiplied with $\nu$ and continuum normalized synthetic observations in a given phase. We search for the value of $\nu$ that minimizes the cross-correlation of residual with stellar spectra and residual with telluric spectra (see Eq. \ref{renorm2}):

\begin{align} \label{renorm}
& r(\nu, \lambda, t) = \Bigg\{ \nu(t) \cdot S_n(\lambda, t, -v_{s}) - \\ \nonumber
& \quad \Big[ \big( F(\lambda) - i_p(\lambda, \phi) \cdot R_p^2/R_{\star}^2 \big) \cdot T(\lambda, t, -v_{s}) \Big] \textstyle \bigotimes \Gamma \Bigg\} \Big/ ( i_a(\lambda, \phi) \textstyle \bigotimes \Gamma ) \\ \nonumber
& \\ 
& \big[ r(\nu) \star F \big] \cdot \big[ r(\nu) \star T \big] = min. \label{renorm2}
\end{align}
There are many ways to formulate an inverse problem for $P$. The form given by Eq. \ref{eqq4} has the advantage of avoiding division by quantities that may turn to zero (telluric transmittance) and of differences between subtracted terms  not being orders of magnitude. With this set-up, the noise may still introduce large fluctuations around the mean value, which now can be controlled. This is done by selecting weights (see below) along the wavelength grid and adding a regularization term ($R$), which controls the balance between the smoothness of the solution and the fit to the observation. Now the minimization problem can be written as
\begin{equation}  \label{eq5}
\begin{array}{c}
\Omega \equiv \Phi + \Lambda \cdot R = min \\
\\
R = \displaystyle \sum_{\lambda} \left( \frac{d P(\lambda)}{d \lambda } \right)^{2}
\end{array}
\end{equation}
The minimum of $\Omega$ is obtained by requiring the derivatives of $\Omega$ to be zero and by linearizing the equation for $R$. Then, $P$ is given by a system of linear equations with a tridiagonal matrix, which is fast and easy to solve. In this exact formulation of Eq. \ref{eqq4}, the telluric absorption acts as a weight; in places with strong absorption ($T\approx0$), the regularization dominates, keeping the oscillations of the solution under control across the affected intervals. The constant $\Lambda$ is a free parameter, which balances the relative importance of regularization. For a given dataset, a value of $\Lambda$ must be chosen and possibly be significantly affecting the recovered solution. In our tests of the method, we can compare the obtained solution to the original "true" exoplanet transmission spectrum (what we put into simulated observations). By searching for the $\Lambda$ that gives the smallest least squares difference between these two, we can find an optimal $\Lambda$. When applying this method to the real observations (with the exoplanet spectrum being unknown), the optimal value of $\Lambda$ can be derived in similar numerical experiments using the correct parameters of the system (size of planet to star, duration of transit, instrumental parameters, estimated S/N). The regularization imposes extra smoothness on the solution, but while this dumps numerical noise, it also reduces spectral resolution of the recovered spectrum in places of high telluric absorption.

An additional advantage of this approach to recovering the exoplanetary spectrum is that one can combine observations of several different transits. As long as each single exposure is combined with corresponding telluric and stellar spectra within the summation in Eq. \ref{eqq4}, adding exposures from several transits is possible without any additional complications. For this we assume that the exoplanet transmittance does not change from one transit to the next.

In this set-up of the inverse problem, stellar flux spectrum and specific intensity spectra at each $\mu$ point, telluric transmittance and relative areas of star and planet are all needed. How to achieve this is covered in Sec. \ref{secReq}, but for now we assume that all these data are available with sufficient accuracy.


\section{Application} \label{Application} 
Now we proceed with testing the proposed method and estimating requirements for both the observed exoplanet system and the instrument. 

        
\subsection{Data sources} \label{Data sources}
For simulating observations and extracting the exoplanet transmission spectrum, three components are needed, stellar spectra (flux and specific intensity), exoplanet atmospheric transmission spectrum, and telluric transmittance. For the tests below, the high-resolution flux and specific intensity spectra of the stars were computed with the stellar atmosphere modelling code MARCS \citep{Gustafsson2008}. One-dimensional hydrostatic MARCS models are based on the most advanced atomic and molecular opacities, and they are extensively tested against real data for the Sun, cool dwarfs, sub-giants, and giants. The quality of the stellar model atmospheres was assessed using spectral synthesis, which also gives the estimate of the quality and completeness of the line lists. In the case of the Sun, the comparison was also done for a wavelength-dependent limb darkening. For simulations of observations and for the data-analysis method, we need the intensity corresponding to the area of the star occulted by the planet or its atmosphere. This is obtained by integrating the specific intensity over the area covered by the planet or by the the semi-transparent atmosphere. Variations in specific intensity across the projection of the planetary disk are much less than 1\% (ignoring the presence of active regions), and the integration over the planetary disk can be replaced safely by the cental value. However, in a some cases this is important to include, such as when observing large planets transiting small stars (especially close to the limb of the star), when only part of the planet covers the stellar disk (being able to include these cases enables us to acquire a few additional exposures at the beginning and end of the transit) or when the exposure time is long and the planet moves across the stellar disk during the exposure. In our tests we have focused on two different cases: a Jupiter-size planet in front of a solar-like star and a super-Earth in front of an M dwarf. Flux and intensity spectra were computed for two stars with parameters according to Table \ref{stellarpar}. 

\begin{table}[h]
\caption{Stellar parameters of the host stars used for simulating transit observations.}
\centering
\begin{tabular}{lll}
\hline \hline
  & solar-like     & M dwarf \\
\hline
Spectral type & G2 & M5 \\
$\mathrm{T_{eff}}$ (K) & 5777 &  3100 \\
log(g) (cgs) & 4.44 & 5.0 \\
$\mathrm{R / R_{\odot}} $& 1.0 & 0.2 \\
\hline
\end{tabular}
\label{stellarpar}
\end{table}
The telluric transmittance was calculated using the atmospheric modelling code LinePak \citep{Gordley1994} implemented on the website http://spectralcalc.com, which provides calculations of transmission spectra of the telluric atmosphere under given conditions. The atmospheric model used in LinePak for integrating the transmission spectrum are the U.S. Standard Atmosphere, which provides pressure and temperature profiles, as well as partial pressures for various gases. In particular, the molecules dominating the NIR absorption such as O$_2$, O$_3$, H$_2$0, CO$_2$, CH$_4$, and HNO$_3$ are all included in the calculations. The line lists used for calculating the absorption bands of these molecules were taken from HITRAN \citep{Rothman2009}. To represent the typical viewing conditions for astronomical instruments, the altitude was set to 2.6 km and the water vapour concentration reduced to 20\% of the default concentration in the U.S. Standard Atmosphere where PWV = 14.3 mm. Observations are then synthesized with an airmass varying from 1.0 to 1.2 during a single transit, to represent the star moving downwards from zenith during three hours of observations.

The projected exoplanetary atmosphere is characterize by its height. We define atmosphere height (often referred to as eclipsing height) as the additional radius that a completely opaque disk would have. This disk blocks the same amount of stellar light as the planet with a semitransparent atmosphere. At wavelengths where the atmosphere is mostly opaque, a larger opaque disk is needed to block the same amount of light, and thus the eclipsing height of the atmosphere is larger at wavelengths where the atmosphere absorbs more light. For the Earth viewed in NIR, the radius would vary between 6~385 km (in wavelength regions with no absorption) and 6~470 km (in wavelength regions with much absorption), making the eclipsing atmosphere height between 0 and 85 km.

\begin{figure}
   \centering
   \includegraphics[width=\hsize]{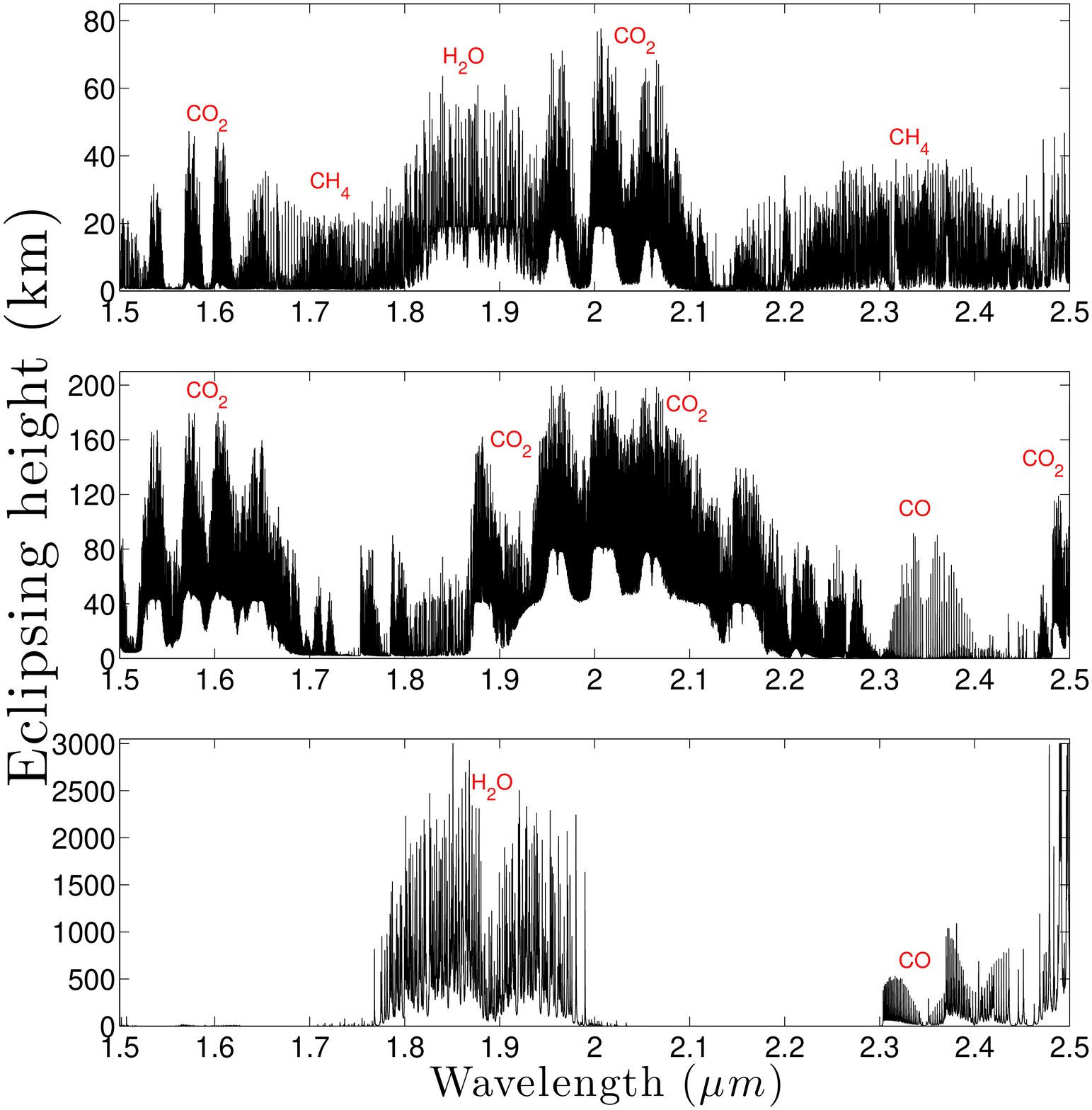}
      \caption{Eclipsing height of model exoplanet atmospheres as function of wavelength in the H and K NIR bands.\newline 
      Upper panel: super-Earth with Earth-like composition.\newline 
      Middle panel: super-Earth with Venus-like composition. \newline 
      Bottom panel: hot-Jupiter.}
         \label{exoplanettransmission}
   \end{figure}

Transmission spectra for super-Earth atmospheres (Fig. \ref{exoplanettransmission}) were created using the same model as the telluric transmission spectrum (LinePak), but increasing the impact parameter of the rays up to 100 km above the surface. We used two different atmospheric models: one with temperature, pressure, and composition identical to Earth and one more Venus-like, with CH$_4$ and H$_2$O replaced by CO and CO$_2$. Using an exo-atmosphere with a different composition than the Earth's enabled us to test that the data analysis method produces artificial features originating in the telluric atmosphere, while an Earth-like atmosphere enabled us to test whether we can reproduce spectral features that also exist in the telluric atmosphere. Super-Earths typically have surface gravity within 0.9 - 1.1 of the Earth's (exoplanet orbit database \citet{Wright2011}), which makes the atmospheric scale height for super-Earths with the same atmospheric temperature and mean molecular weight as the Earth or very similar to the Earth's (8.5 km). Therefore we used the atmosphere height directly from the telluric model. For the Venus-like super-Earth, we wanted to test a more extended atmosphere. If the planet is hotter, the scale height will increase. We used a scale height that is twice as large as for the Earth. On top of this, the CO$_2$ in the atmosphere absorbs light very efficiently, making the eclipsing atmosphere height of the Venus-like planet almost 200 km. For planetary radius, we used 2.5 $\mathrm{R_\oplus}$ in both cases. 

Hot-Jupiter model spectra (Fig. \ref{exoplanettransmission}) were derived in a similar manner, computing the optical length of the rays at different distances from planet centre. The total transmittance was then evaluated using the projected areas of the corresponding concentric rings. The atmospheric structure was taken from \citet{Miguel2014ApJ}. Opacity contributors used in this model include H$_2$, CH$_4$, CO, N$_2$, NH$_3$, H$_2$O, PH$_3$, H$_2$S, TiO, VO, CaH, MgH, FeH, CrH, SiO, CaOH, and TiH. (We do not include any clouds or hazes.) The model atmosphere height is 5~600 km (scale height 500 km) on top of a completely opaque core with radius 90~000 km. This model planet is slightly larger than the average hot-Jupiter. We scale the radius and eclipsing atmosphere height to match a Jupiter-size planet (radius 69~900 km). Roughly four-fifths of currently known hot-Jupiters are Jupiter size or larger, making this a conservative estimate for the size of a hot-Jupiter.

\subsection{Simulation set-up} \label{Simulation set-up}
We simulated observations of three cases (hot-Jupiter transiting solar-like star and Venus and Earth-like rocky planets transiting M dwarfs) during transit and applied the data-analysis method to these "observations". Simulations were performed with a hypothetical state of the art high-resolution spectrograph with spectral resolution R=100~000 and simultaneous wavelength coverage 1.5 - 2.5~$\mu m$. The S/N of simulated observations is based on estimates of what modern instruments at 8m class telescopes will be able to achieve during 75 s exposures of the two stars. For a solar-like star (K mag 8) S/N $\approx$ 150 and M-dwarf (K mag 9) S/N $\approx$ 90 per resolution element. The S/N is estimated to be at the centre of the K band. For simplicity we assumed that the S/N is constant throughout the observed spectral range and ignored readout and sky noise (sky emission discussed in Sec. \ref{subsecMetReq}). If these noise sources are significant, the integration time should be adjusted to reach desirable S/N levels. 

\begin{figure}
\centering
\includegraphics[width=\hsize]{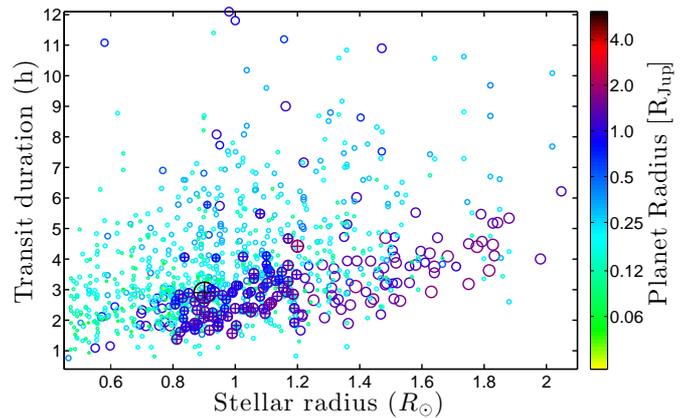}
\caption{Transit duration versus stellar radius for 4500 Kepler objects. \newline
Colour coding and size of symbols indicate the planetary radius. The $\mathrm \oplus$ symbols denote hot-Jupiters orbiting solar-like stars.}
\label{Transitduration}
\end{figure}

The duration of a transit is an important parameter when estimating the capabilities of the data analysis method, since it directly limits the total observing time. Transit durations range from slightly below one to many hours, approaching a whole day in extreme cases. The correlation between the size of the star and transit duration is not straightforward. Planets transiting larger stars cross a larger stellar disk, but at the same time the stars are more massive, meaning higher orbital velocities. On top of this, the semi major axis (and to a lesser degree the inclination and eccentricity of the orbit) affects the transit duration. Looking at statistics of transit durations from the the Exoplanet Orbit Database \citep{Wright2011}, we find longer average transit duration for planets orbiting larger stars (see Fig. \ref{Transitduration}). For hot-Jupiters transiting solar-like stars (here taken as planets with $R_p > 0.5 R_{Jup}$, $a < 0.3 au$ and orbiting stars with $0.8 R_{\odot} < R_{\star} > 1.2 R_{\odot}$), the median transit duration is 165 minutes. We use this as the transit duration for the hot-Jupiter case. 

For the Earth-like super-Earth, we place this planet in the habitable zone (the distance from the star at which liquid water is possible). The inner and outer edges of the habitable zone depend on many things, such as the planet's surface gravity, atmospheric pressure, and greenhouse effects. The eccentricity of the orbit and the rotation of the planet complicate this further. To get rough estimates of the habitable zone, we use the orbital distance of the Earth and scale it with inverse square of stellar luminosity, and this distance is then converted to a transit duration (assuming inclination $90^{\circ}$. For an M dwarf this would give a transit duration of 90 minutes. For the Venus-like super-Earth, we use the same method and scale the semi-major axis to match the semi-major axis of Venus, giving us a transit duration of 75 minutes. See Table \ref{planetpar} for details on orbital parameters.

\begin{figure}
\centering
\includegraphics[width=\hsize]{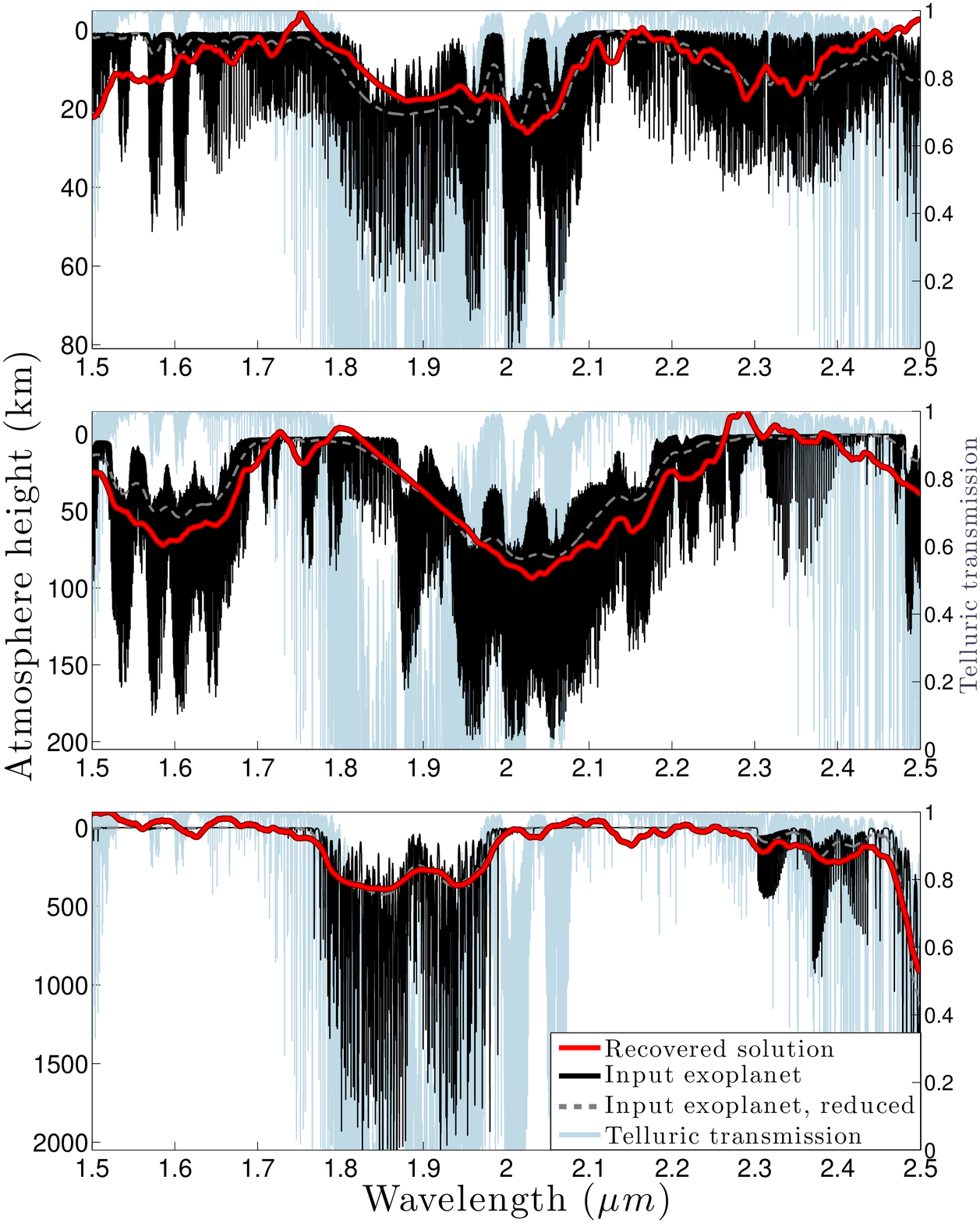}
\caption{Recovered exoplanetary transmission spectra from simulated transit observations. \newline
Upper panel: Earth-like super-Earth transiting M5 star, 4 transits;\newline
Middle panel: Venus-like super-Earth transiting M5 star, 1 transit;\newline
Lower panel: hot-Jupiter transiting G2 star, 1 transit.}
\label{G2M4M4}
\end{figure}

For each case we show four spectra in the same panel (Fig. \ref{G2M4M4}). The solution of the inverse problem, corresponding to the recovered exoplanet transmission spectrum, is shown in red. To show how accurate the recovery is, the original transmission spectrum of the exoplanetary atmosphere is shown in black. Along with this we over plot (dashed grey line) the input exoplanet spectrum with the same amount of weighted regularization as the recovered solution. These spectra are represented by their atmospheric height (defined as in Sec. \ref{Data sources}), with the axis on left-hand side of each figure. We also include the telluric transmission in greyish blue to indicate the parts of the wavelength regime where the telluric transmittance, hence the signal, is very low. This is represented by the transmittance axis on the right-hand side of each figure.

Results from the tests are shown in Fig. \ref{G2M4M4}. For the hot-Jupiter ($R = 1 R_{Jup}$) transiting a solar-like star, the strongest recovered absorption bands coincide with the major absorption bands in the input spectrum. In the tests with a super-Earth with Earth-like transmittance, we had to combine observations from four transits (corresponding to a total of 6 hours of observations) in order to recover the major H$_2$O, CO$_2$, and CH$_4$ bands, and even with observations from four transits, the results are noisy and contain a few spurious features. The Venus-like exoplanet is more advantageous than the Earth-like case mainly because of its larger atmosphere, and we manage to recover the major absorption bands during a single 75-minute-long transit.

\begin{table}[h]
\caption{Physical properties of star-exoplanet systems.}
\centering
\begin{tabular}{llll}
\hline \hline
&Hot&Earth-&Venus-\\
&Jupiter&like&like\\
\hline
Semi-major axis (au) &  0.054 & 0.064 & 0.046\\
Inclination $i$ & $90^{\circ}$ & $90^{\circ}$ & $90^{\circ}$\\
Orbital period (days) & 4.5 & 13.3 & 8.1\\
Transit duration (min) & 165 & 90 & 75\\
$\mathrm R_p^2/R_{\star}^2$ & 0.010 & 0.013 & 0.013\\
System velocity $v_{s}$ (km/s) & 15 & 15 & 15\\
Orbital velocity $K_{p}$ (km/s) & 130 & 52 & 62\\
Velocity semi-amplitude $K_{\star}$ (m/s) & 124 & 6.5 & 7.0\\
K mag & 8.0 & 9.0 & 9.0\\
\hline
\end{tabular}
\label{planetpar}
\end{table}

\section{Requirements} \label{secReq}
Owing to the very small fraction of the stellar disk that is blocked by the exoplanet atmosphere, recovering exoplanetary transmission spectra puts heavy requirements on observations, input data, and the targeted planetary system. Based on simulations of varying input parameter, we tried to estimate requirements for transit spectroscopy using this data analysis method. Unless noted otherwise, all tests in this section are done with the same planet and star parameters as the Venus-like super-Earth case above, see Table \ref{planetpar}.

\subsection{Observational requirements} \label{Observational requirements}
The two most important characteristics of observations, and therefore the instrument, are (1) spectral resolution and (2) wavelength coverage.

\noindent 1. Spectral resolution: The resolution of the spectrograph is a vital aspect of the instrument. With high resolution, we are able to see between telluric lines and detect contribution from the exoplanet that might not be observable if using lower resolution. We can use this to see molecular species present in both the telluric and the exoplanetary atmosphere, since there will be a small radial velocity difference between the telluric atmosphere and the exoplanet. This will shift the exoplanetary lines relative to the telluric lines, and this difference is detectable with
high resolution. With a varying radial velocity difference during a transit, we will also be able to scan for exoplanetary lines superpositioned on telluric features,
thanks to the rotation of the Earth and the orbit of the exoplanet. 

\begin{figure}
\centering
\includegraphics[width=\hsize]{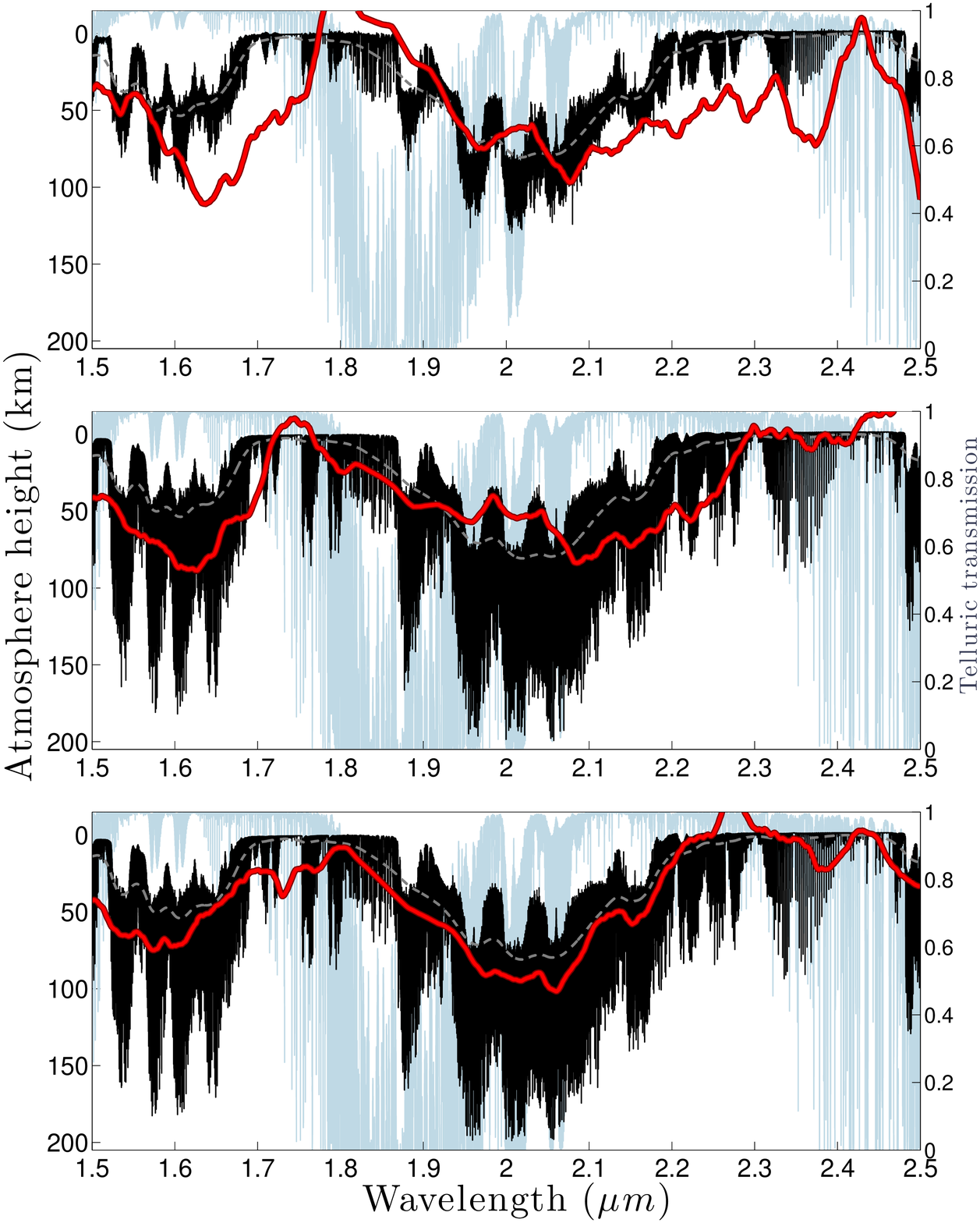}
\caption{Recovered exoplanetary transmission spectra from simulated transit observations of hot super-Earth transiting M5 star. We simulate observations using instruments with different spectral resolution to show the importance of high spectral resolution. \newline
Upper panel: spectral resolution: R=10~000; \newline
Middle panel: spectral resolution: R=50~000; \newline
Lower panel: spectral resolution: R=100~000. \newline
Colour coding is the same as in Fig. \ref{G2M4M4}.}
\label{Low resolution}
\end{figure}

To show the importance of high resolution, we simulated observations with spectral resolution of 10~000, 50~000 and 100~000 and applied the data-analysis method. Results of these tests are shown in Fig. \ref{Low resolution}. Recovering the exoplanet spectrum is possible at resolutions less than 100~000, but better spectral resolution relaxes the required S/N. At resolution R = 50~000 recovering exoplanet spectrum in regions with strong telluric absorption is difficult. At spectral resolution below 50~000, telluric and exoatmospheric lines blend, and the recovery of exoplanet spectrum deteriorates. It is no longer possible to differentiate spurious features from true absorption bands.

\begin{figure*}
\includegraphics[width=\hsize]{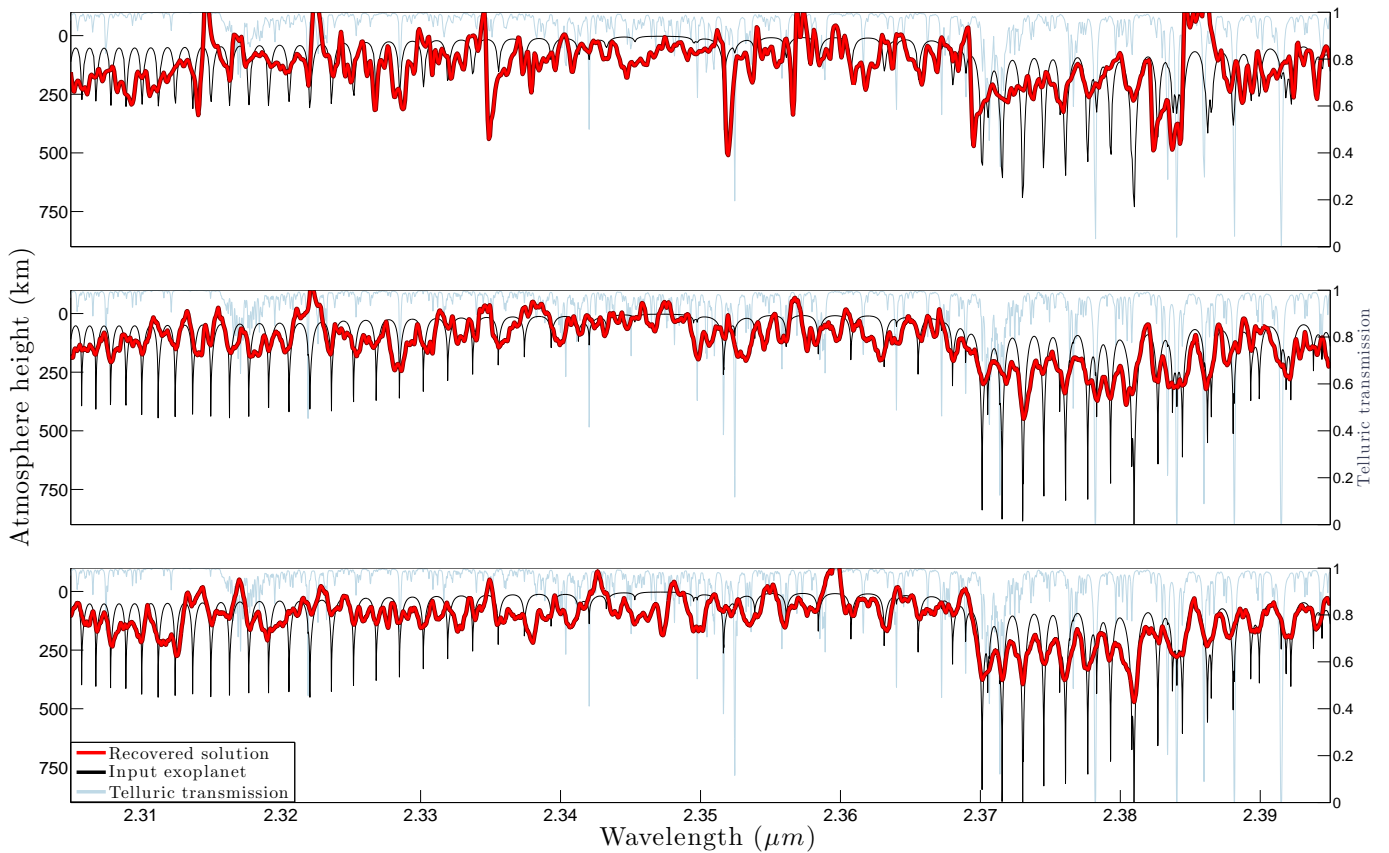}
\caption{Recovered exoplanetary transmission spectra from simulated transit observations of a hot-Jupiter transiting G2 star, combing observations of 15 transits. We demonstrate the reconstruction of individual absorption lines made possible by high S/N and spectral resolution. \newline
Upper panel: Spectral resolution: R=10~000; \newline
Middle panel: Spectral resolution: R=50~000; \newline
Lower panel: Spectral resolution: R=100~000.}
\label{Zoom in}
\end{figure*}

There are two factors that affect the resolution of the recovered exoplanet spectrum, the instrumental resolution and the regularization parameter. If data quality is very high (i.e. high S/N and/or many combined transits), one can afford to reduce regularization and still recover exoplanet transmission spectrum with high accuracy. In the most favourable cases, when combining many transits of bright star with a transiting hot-Jupiter, we are able to reconstruct individual lines. To illustrate the effect of high resolution, we simulated observations from 15 transits of the same Jupiter size exoplanet and solar-like star as used previously. We do this with spectral resolutions of 100~000, 50~000, and 10~000. We show a small region of the reconstructed exoplanetary spectrum where we reconstruct individual lines (see Fig. \ref{Zoom in}). When using spectral resolution of 100~000 and 50~000 the spectral features (CO lines in this case) are reconstructed. When using a resolution of 10~000, the results are significantly worse, and it becomes hard to distinguish between real and spurious features. While the recovered solution appears to have the same spectral resolution (owing to similar amounts of regularization), the difference in the accuracy of the reconstruction depends only here on the resolution of the spectrograph (and thus on the ability to differentiate exoplanetary contribution from telluric lines).

Since the telluric transmittance in NIR contains many spectral features, it is critical to know the telluric atmosphere to high accuracy. In Sec. \ref{subsecMetReq} we describe why high resolution can be useful when obtaining telluric transmission spectra.  

\noindent 2. The simultaneous wavelength coverage of the spectrograph is another important factor. Being able to detect a continuum and absorption features simultaneously enables us to correct for errors in planetary radius or estimated relative brightness decrease, which will shift the recovered spectrum along the eclipsing height axis. At least 0.1 $\mu m$ is needed to target the most favourable bands. With very high S/N, detecting a continuum is possible even in narrow bands (such as in Fig. \ref{Zoom in}). 

Being able to cover a significantly wider wavelength range than 0.1 $\mu m$ in a single exposure is very valuable. Observing time is limited by the transit duration, and wide wavelength coverage makes it possible to search for several species simultaneously, as well as to identify what species is responsible for an absorbing region from the shape and component spacing on the feature.

Broader spectral coverage also helps with the continuum normalization of observations. The quality of this step in stellar spectroscopy depends very much on the spectral type of the target and the spectral domain. For cooler stars, molecular bands can prevent proper normalization based solely on the science spectrum. An alternative approach is to measure relative monochromatic throughput of telescope + spectrometer using a well known spectrum. We have tested this techniques with the solar reference spectrum in the optical and NIR using the observations of Vesta and Ganymede taken with the HARPS (ESO La Silla) and the HIRES (Keck) spectrometers. Previous knowledge of the result provided the throughput curves, thereby correcting all small-scale instrumental and detector inhomogeneities to better than 0.5\%. Then when applied to the science target, the remaining trend is only due to the difference in spectral type and could be easily adjusted for.

We have chosen to focus on the NIR wavelength regime in this paper. The main benefit to this is the numerous molecular absorption bands located there. In hot-Jupiter atmospheres, we expect to find species, such as H$_2$O, CO$_2$, CH$_4$, CO, H, H$_2$, O, and OH \citep{Miguel2014ApJ}. The expected chemical composition of super-Earths is more uncertain. Extrapolating from the atmospheres of rocky planets in out own solar system indicates that they are likely to be very diverse. The H, K, L, and M bands in NIR contain many molecular absorption bands from important molecules, such as O$_3$, CO, CO$_2$, SO$_2$, CH$_4$, and NH$_3$. In wavelengths longer than 2.5 $\mu m$, sky emission becomes significant, making ground-based observations less competitive with space observations. At shorter wavelengths, sky emission is less significant. In Sec. \ref{subsecMetReq} we propose dealing with sky emission using spectral synthesis. The S/N required to recover the exoplanet transmission spectrum is high, so using wavelengths longer than 3 $\mu m$ will be problematic for all but the largest planets and brightest stars. We are therefore limited to the H and K-bands (1.5 - 1.8 $\mu m$ and 1.9 - 2.5 $\mu m$), where we have strong absorption from CO$_2$, CO, CH$_4$, H$_2$O, CH$_4$, and NH$_3$. Observations in the J-band (1.1 - 1.4 $\mu m$) can also be valuable due to the strong O$_2$ absorption band around 1.27 $\mu m$, but unfortunately very few other molecules have absorption bands in this region.

\subsection{Model requirements} \label{subsecMetReq}
To test the method's robustness, we examined how sensitive our method is to the accuracy of measurements and model spectra.
We need telluric transmission spectra separately from the stellar contribution. Observations consists of several short exposures of the system as the planet moves across the stellar disk, and for each of these exposures, the matching telluric transmittance is needed. This must be very accurate and include changes in the telluric atmosphere during the transit. This could be due to changes in airmass as the star moves across the sky or it could be due to changes in the relative abundances of specific species (most notably water vapour). 

Such changes will affect line strength and saturation, and small errors in the telluric transmittance used in the data analysis can introduce spurious features in the recovered exoplanetary spectrum. For this reason we do not recommend using pre- and post-transit observations of telluric standards for obtaining telluric transmittance, since it will be insensitive to changes in the atmosphere during the transit. Dedicating time observing a telluric standard at regular intervals during the transit is not recommended either, because the standard must lie close to the target where small changes in telluric transmittance will affect results. 

Since a transit event is limited in time, we want to avoid spending time during the transit not observing the system itself. A more promising method of obtaining the telluric transmittance is synthetic spectra fitted to observations. Several tools for synthesizing high-precision telluric transmission spectra exist, are under development \citep{Seifahrt2010A&A, Kausch2014ASPC}, and can currently reproduce telluric absorption to an accuracy better than methods using telluric standard stars. Ideally such spectral synthesizing includes sky emission to reduce the need for nodding, thus maximizing the time spent observing the transiting planet. In spectral regions where sky emission is significant, nodding might be required in order to produce a telluric model with high enough accuracy. In such cases observation time should take this into account, in our estimations we assume sky emission is either negligible or can be modelled with high accuracy.

To minimize the risk of including contributions from the exoplanet when fitting telluric transmittance to observations, we need high enough spectral resolution to resolve the Doppler shift due to differences in radial velocity between telluric and exoplanet atmosphere. Observations will originally be in the rest frame of the detector. To avoid introducing additional errors, telluric model spectra (and stellar models) should be fitted to observations at this stage, before they are shifted and interpolated onto rest frame of the exoplanet. By using synthetic telluric spectra, we do not need to spend any of the limited observing time during the transit to observe telluric spectrum template, but can instead spend all available time to observe the transiting exoplanet. Observations aimed at measuring the telluric atmosphere, by observing telluric standard stars or similar objects with known spectral features along the same line of sight as the transiting exoplanet, could still be carried out before and after the transit in order to get the starting and ending conditions of the atmosphere.

\begin{figure}
\centering
\includegraphics[width=\hsize]{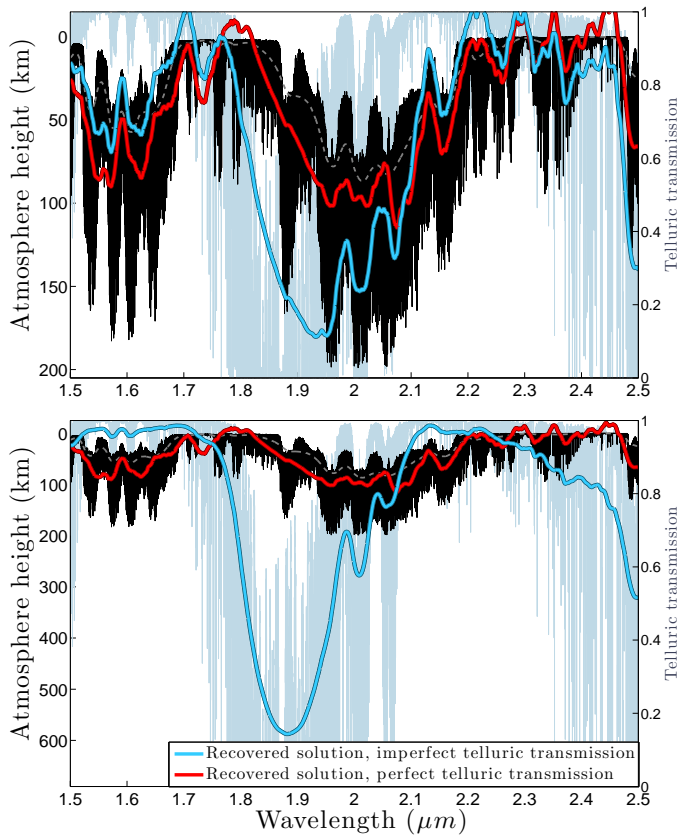}
\caption{Recovered exoplanetary transmission spectra from simulated transit observations of a hot super-Earth transiting M5 star. Here we test the effects of using imperfect telluric transmittance in the data analysis. \newline
Upper panel: Random over/underestimation up to $\pm$2\% of telluric transmittance; \newline
Lower panel: Systematic underestimation up to $\pm$2\% of telluric transmittance. \newline
Colour coding is the same as in Fig. \ref{G2M4M4}.}
\label{Imperfect telluric}
\end{figure}

The accuracy of the telluric spectra must be high; systematic errors such as constant over- or under-estimation of telluric features manifests as spurious features in the solution. To investigate the needed level of accuracy, we tested cases with imperfect telluric spectra when applying the data-analysis method. First we used small non-systematic errors by alternating between over- and under-estimation of the telluric transmittance by up to $\pm$2\%. We still manage to recover the same major molecular absorption bands as in the case-perfect telluric transmittance, but artificial features were producedt in regions with strong telluric absorption. The effects of systematic errors were also examined, by always underestimating the telluric transmittance by up to 2\% when performing the data analysis. In this case, the reconstruction is completely dominated by the errors in telluric transmittance, and the contribution from the exoplanet is no longer visible. Both tests are shown in Fig.~\ref{Imperfect telluric}.

\begin{figure}
\centering
\includegraphics[width=\hsize]{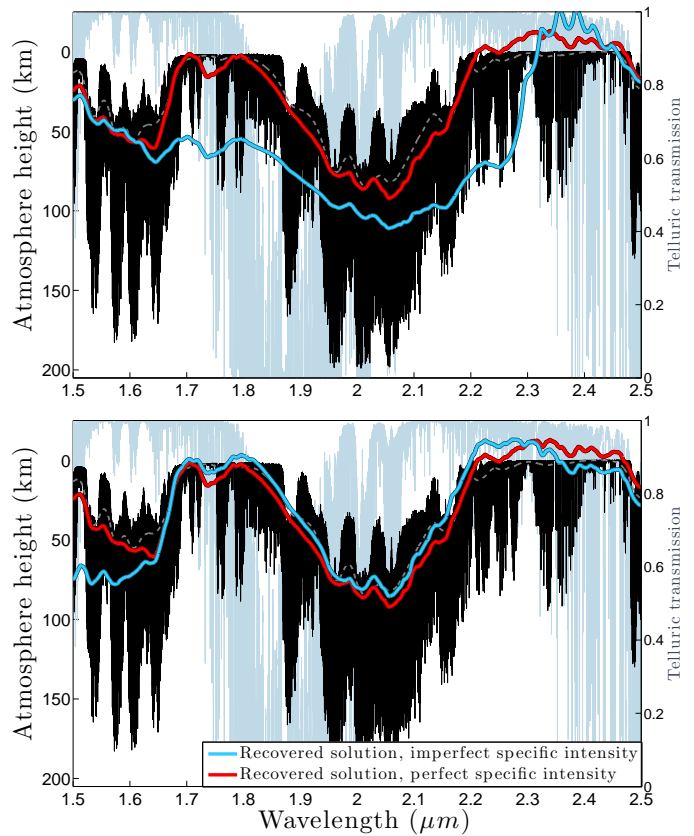} 
\caption{Recovered exoplanetary transmission spectra from simulated transit observations of a hot super-Earth transiting M5 star. Here we test the effects of using imperfect stellar specific intensity spectra in the data analysis. \newline
Upper panel: Specific intensity from quadratic limb darkening equations; \newline
Lower panel: Specific intensity from model atmosphere with 5\% too high effective temperature.\newline
Colour coding is the same as in Fig. \ref{G2M4M4}.}
\label{Imperfect intensity}
\end{figure}

The flux spectrum of the host star is needed for the inversion. This can be measured before or after the transit. Its stability or variability can be monitored outside the transit. Specific intensity spectra is more challenging. One approach is to use synthetic spectra based on a custom-made model of stellar atmosphere. The model can be validated by comparing the flux spectrum with observations outside transit. Limb darkening dependence is hard to verify. We examine the effects of errors in specific intensity by using imperfect specific intensity spectra in the data analysis. We use quadratic limb darkening equation in one test and modelled specific intensities with effective temperature 5\% higher than for the synthesis of observations. Results can be seen in Fig. \ref{Imperfect intensity}. In both cases, the major absorption bands from the exoplanet are still recovered, but with limb darkening equations, errors in the shape of the continuum are introduced, while with imperfect model spectra, we only observe minor deviations from perfect intensity spectra. It is worth considering that the case with limb darkening equations is close to optimal, while the case with model spectra  introduces deliberate errors. Model spectra still produce better results, so we recommend the use of model spectra over limb darkening equations. 

Approximating the area-integrated specific intensity ($i_p$ and $i_a$ in Eqs. \ref{eqq1}) with the value at the centre of the planet or using incorrect $\mu$-values will produce effects similar to the test above. In the cases presented here (hot-Jupiter in front of solar size star and super-Earth in front of M dwarf), the errors in specific intensity spectra will be less than 1\%, such small errors will not affect the end results in any meaningful way.

Inhomogeneities on the stellar surface, such as starspots, can potentially also cause artificial features in the recovered solution. If the transiting exoplanet crosses one or several starspots, the light passing through the atmosphere will be significantly different from the assumed specific intensity used in the data analysis. The effects this has on the recovered exoplanet spectrum depends on the amount of exposures that are contaminated and on the fraction of the exoatmosphere that covers the spots. For young active stars, a significant fraction of the stellar surface can be covered by starspots. Transit spectroscopy will be very complicated for these stars. For solar-like stars, the amount of contamination from starspots will very likely be insignificant. For the small fully convective stars, starspot patterns are largely unknown and could cause problems depending of the size of individual spots and the fraction of the disk that is covered. 

\begin{figure}
\centering
\includegraphics[width=\hsize]{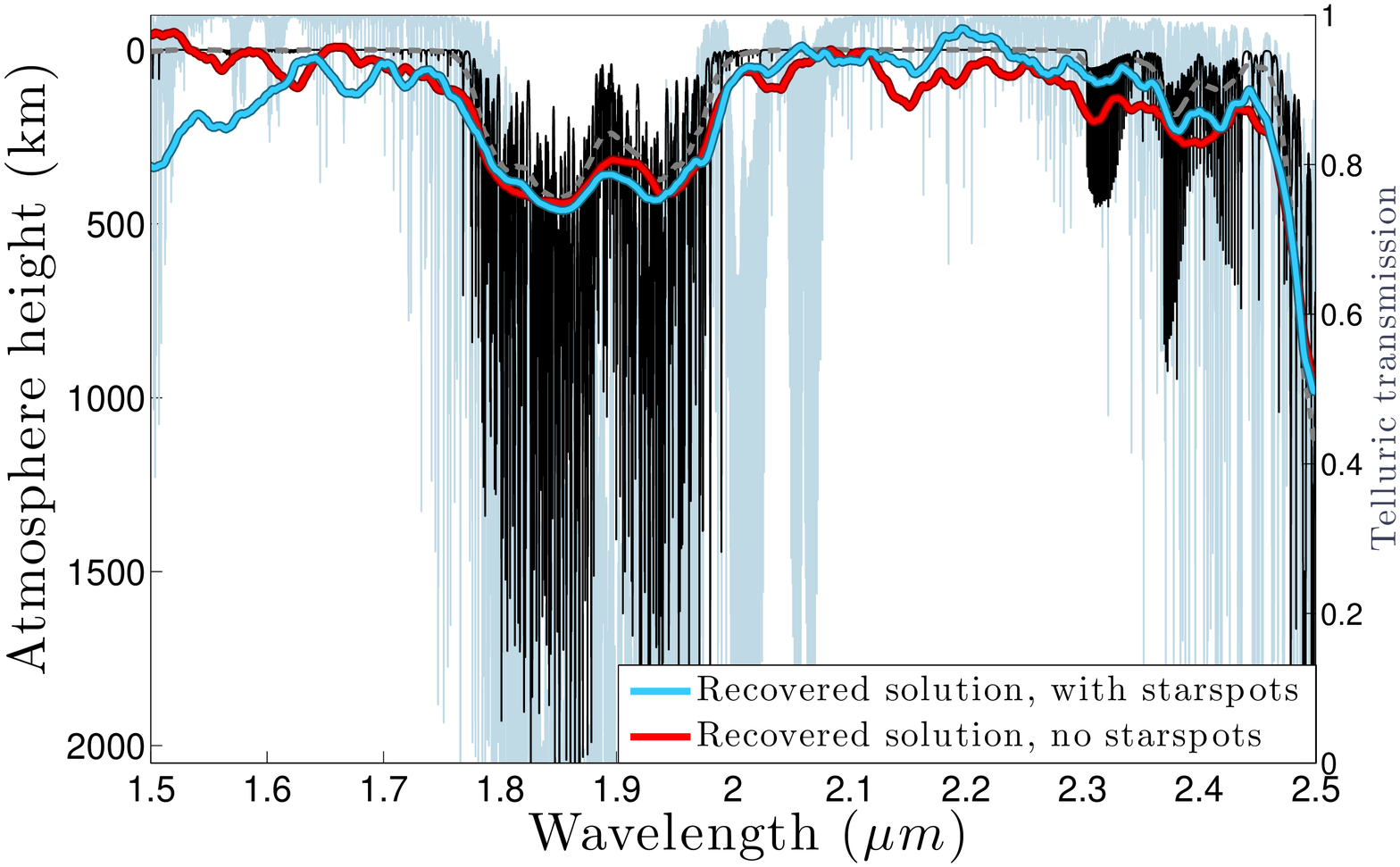}
\caption{Recovered exoplanetary transmission spectra from simulated transit observations of a hot-Jupiter transiting active G2 star. Here the planet crosses starspots during observations, this is then ignored in the data analysis. \newline
Colour coding is the same as in Fig. \ref{G2M4M4}.}
\label{Starspots}
\end{figure}

We performed a simple test where we simulate observations of a transiting hot-Jupiter crossing a solar-like star, where we replace some of the s	pecific intensity spectra with the spectra of sunspots (data from Pierce Solar Telescope). We then ignore the spots when applying the data analysis method. For this test we used spots of sizes comparable to the planet, and a rather extreme case where the planet covers spots in 20\% of the exposures (>50\% of planet covered in 10\% of exposures). Results from this test can be seen in Fig. \ref{Starspots}. We managed to recover the same spectral features as when using a starspot free case. We therefore do not expect any major issues concerning starspots for solar-like stars. Ideally exposures contaminated by contribution from starspots should be removed from the data. The photometry of the transits can detect the starspots if the spots are large enough to produce a detectable brightness increase in the transit light curve as the planet passes over it. This spot detection has successfully been demonstrated by \citet{Pont2007A&A, Carter2011, Sing2011MNRAS}. 

Photometric measurements taken simultaneous with the spectroscopic measurements could help find the most problematic spots so that they can be removed from the data, at the same time giving the exact start and end of the transit. Unocculted starspots and other forms of stellar variability during the transit could potentially also cause problems, mainly in the form of errors in size determination of the solid part of the exoplanet. In Sec. \ref{subsecTarSel} we explore the effects of errors in planetary radius. Stars covered by a large number of star spots, enough to alter the flux significantly, could also be problematic. In such cases the stellar spectrum underneath the planet will be different from the stellar flux, which would make synthetic specific intensity spectra less accurate if using observed stellar flux to validate the stellar model. For stars with solar-like spot coverage (less than 0.5\% of the sun is covered by spots during solar maximum), this effect would be insignificant.

\subsection{Target selection and planet parameters} \label{subsecTarSel}
\begin{figure}
\centering
\includegraphics[width=\hsize]{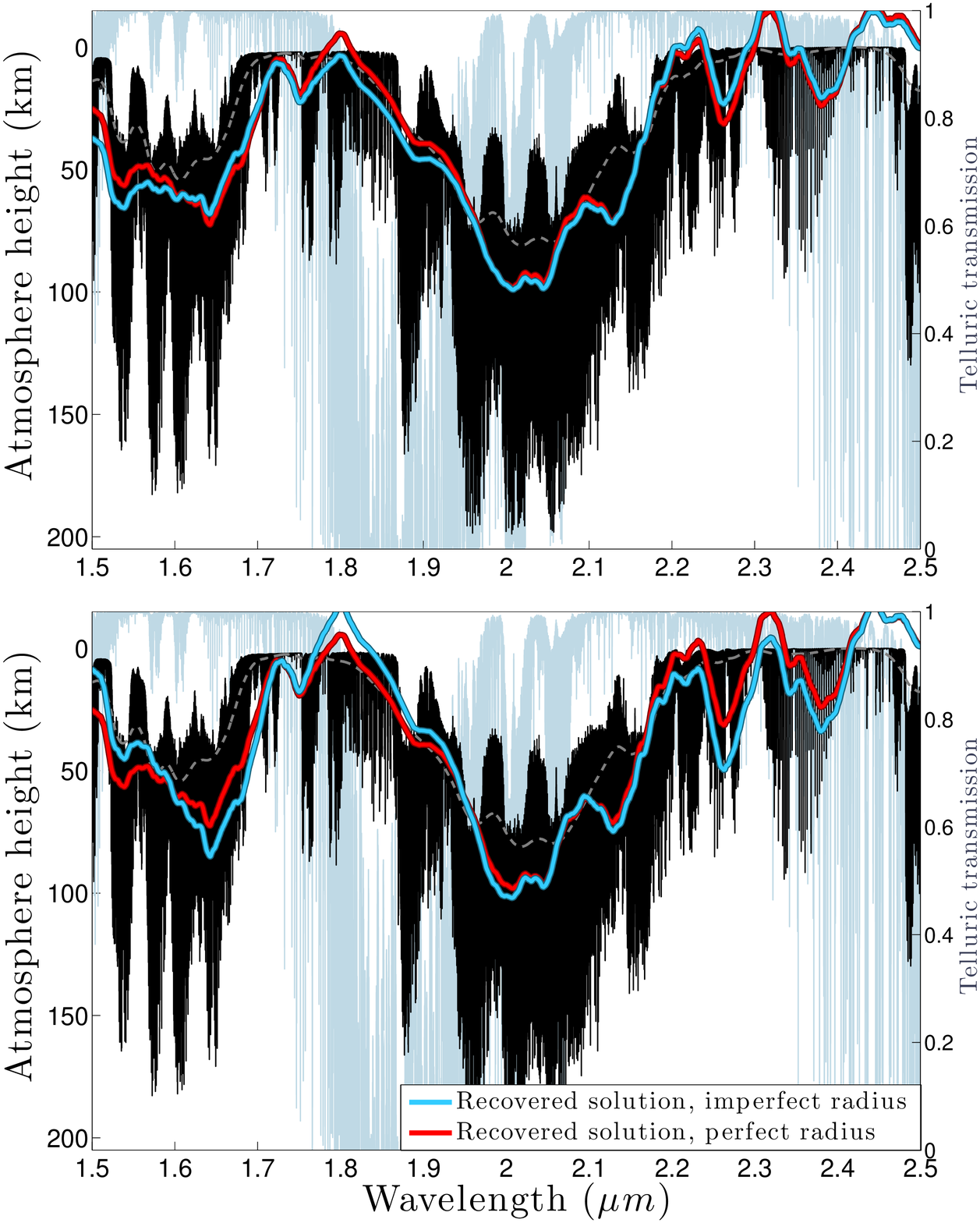}
\caption{Recovered exoplanetary transmission spectra from simulated transit observations of hot super-Earth transiting M5 star. Here we test the effects of using errors in estimated planetary radius in the data analysis. \newline
Upper panel: 10\% over-estimation of planetary radius; \newline
Lower panel: 10\% under-estimation of planetary radius.\newline
Colour coding is the same as in Fig. \ref{G2M4M4}.}
\label{errorRadius}
\end{figure}

The data analysis method requires the size of the planet to be known. This can be measured in the photometric transit light curve. Errors and uncertainties in the size of the planet will have two effects. First, this will shift the recovered spectrum along the atmospheric height axis, down for over-estimations and up for under-estimations. This effect is easily minimized by setting the continuum of the solution to zero. The second effect of errors in the size of the planet is the expected brightness decrease in the star as the planet moves in front of the stellar disk (characterized $\nu$ in Eq. \ref{eqq4}). Errors in this can leave an imprint of stellar or telluric spectra in the recovered solution. This can be solved by using our proposed way of calculating $\nu$ (Eq. \ref{renorm}), which corrects for errors in planetary size. Reasonably small errors in planetary size will only have a weak effect on the relative strength of absorption features. We show this by simulating two cases, one with 10\% overestimation and one with 10\% underestimation of planetary radius, which are the results after setting continuum to zero shown in Fig. \ref{errorRadius}. 

\begin{figure}
\centering
\includegraphics[width=\hsize]{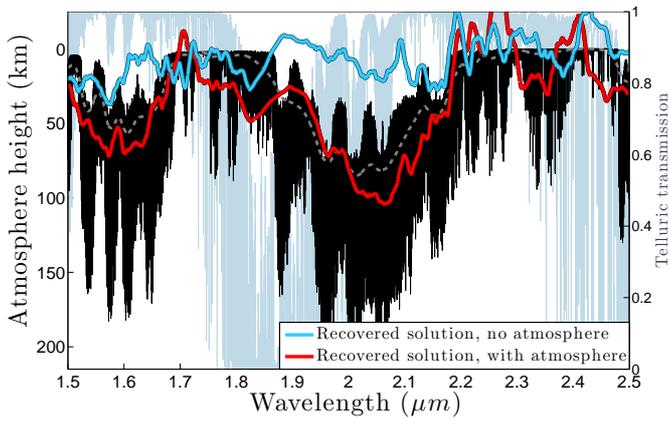}
\caption{Recovered exoplanetary transmission spectra from simulated transit observations of super-Earth transiting M5 star. In one case the exoplanet has an atmosphere, and in the other the exoplanet has no atmosphere, so here we show that this difference can be detected. \newline
Colour coding is the same as in Fig. \ref{G2M4M4}.}
\label{NoAtmo}
\end{figure}

Another potential problem is the case where there is no atmosphere (or a completely transparent or opaque atmosphere in all measured wavelengths). We investigated the ability of the data-analysis method to distinguish between a planet with a semi-transparent atmosphere and a planet with no atmosphere at all by simulation observations of the two cases, and then applied the data analysis to both cases in exactly the same way. Results are shown in Fig. \ref{NoAtmo}. In the case with no atmosphere, the deviations from a flat continuum at height 0 are mainly due to the inverse method trying to fit the noise. In this case we cannot distinguish a planet with no atmosphere from a planet with a thin atmosphere (eclipsing height less than 100 km). This upper limit can be reduced
with less noisy observations (or observations from several transits). An atmosphere dominated by Rayleigh scattering in the observed wavelengths is similar to the case with no atmosphere. The needed level of precision to measure the small changes across the wavelengths range due to Rayleigh scattering is difficult to achieve with ground-based telescopes and could easily be mistaken for a planet with no atmosphere using this method. Here other observational techniques would be more appropriate, such as low-resolution space-based spectroscopy.

\begin{table*}
\caption{Required S/N per resolution element (R=100~000), all exposures combined.}
\centering          
\begin{tabular}{l l l l l }     
\hline\hline       
Star & hot-Jupiter ($1.0\mathrm{R_{Jup}}$) & Hot Neptune ($1.0\mathrm{R_{Nep}}$) & Venus-like super-Earth ($2.5\mathrm{R_\oplus}$) & Earth-twin ($1.0\mathrm{R_\oplus}$)\\
G2-star ($1.0 \mathrm{R_\odot}$) & $\>$ S/N>400 & S/N>3~500 & S/N>20~000 &  S/N>150~000\\
K2-star ($0.8 \mathrm{R_\odot}$) & $\>$ S/N>350 & S/N>2~500 & S/N>12~000 &  S/N>90~000 \\
K8-star ($0.5 \mathrm{R_\odot}$) & $\>$ S/N>100 & S/N>1~000 & S/N>5~000 & S/N>36~000 \\
M5-star ($0.2 \mathrm{R_\odot}$) & $\>$ S/N>60 & S/N>200 &  S/N>800 &  S/N>5~700 \\
\hline           
\end{tabular}
\label{RequiredSN}
\end{table*}

Planetary systems come in many different sizes and brightnesses. The relative projected area of the planet's semitransparent atmosphere to the area of the stellar disk is the most important factor for the required S/N for transit spectroscopy. This can differ by a factor of 10~000 between the most extreme cases. We applied the method to simulations of a multitude of different size ratios. To make it easier to account for different transit durations and observations during several transits, this was converted into S/N per resolution element of all exposures combined (see Table \ref{RequiredSN}). This is for the idealized case and that eclipsing height of the atmosphere strongly affects the required S/N.

The constantly growing body of known exoplanets includes many suitable candidates for transit spectroscopy of this type. In Table \ref{obstargets} we list some the most advantages exoplanets for observations of them. Data was taken from the Exoplanet Orbit Database \citep{Wright2011}.

\begin{table}[h]
\caption{Currently known most suitable observational targets for the observational and data analysis method presented in this paper}.
\centering
\begin{tabular}{lllc}
\hline \hline
Exoplanet &  $\mathrm{R_{planet}}$ & $\mathrm{R_{star}}$ & 2MASS K$_S$\\
\hline
\textbf{hot-Jupiters}\\
\object{HD 209458 b} & $1.4 \mathrm{R_{Jup}}$ & $1.2 \mathrm{R_{\odot}}$ & 6.3 \\
\object{HD 189733 b} & $1.1 \mathrm{R_{Jup}}$ & $0.76 \mathrm{R_{\odot}}$ & 5.5 \\
\object{HD 80606 b} & $1.0 \mathrm{R_{Jup}}$ & $1.0 \mathrm{R_{\odot}}$ & 7.3 \\
\object{WASP-80 b} & $0.95 \mathrm{R_{Jup}}$ & $0.57 \mathrm{R_{\odot}}$ &       8.4\\
\textbf{Hot Neptunes}\\
\object{HAT-P-26 b}      & $1.6 \mathrm{R_{Nep}}$ & $0.79 \mathrm{R_{\odot}}$ &       9.6\\
\object{HAT-P-11 b}      & $1.2 \mathrm{R_{Nep}}$ & $0.75 \mathrm{R_{\odot}}$ & 7.0\\
\object{GJ 436 b} & $1.1 \mathrm{R_{Nep}}$ &     $0.46 \mathrm{R_{\odot}}$ &        6.1\\
\object{GJ 3470 b} & $1.1 \mathrm{R_{Nep}}$ & $0.50 \mathrm{R_{\odot}}$ & 8.0\\
\textbf{super-Earths}\\
\object{GJ 1214 b}       & $2.6 \mathrm{R_{\oplus}}$ & $0.21 \mathrm{R_{\odot}}$ &       8.8\\
\object{Kepler-49 b} & $2.7 \mathrm{R_{\oplus}}$ & $0.53 \mathrm{R_{\odot}}$ &       12.4\\
\object{Kepler-125 b} & $2.3 \mathrm{R_{\oplus}}$ & $0.51 \mathrm{R_{\odot}}$ & 11.7\\
\object{HD 97658 b} & $2.3 \mathrm{R_{\oplus}}$ & $0.70 \mathrm{R_{\odot}}$ & 5.7 \\
\object{55 cnc e} & $2.0 \mathrm{R_{\oplus}}$ & $0.94 \mathrm{R_{\odot}}$ &       4.0\\
\textbf{Earth-size planets}\\
\object{Kepler-205 c} &         $1.6 \mathrm{R_{\oplus}}$ & $0.55 \mathrm{R_{\odot}}$ &       10.8\\
\object{Kepler-186 d} &         $1.4 \mathrm{R_{\oplus}}$ & $0.47 \mathrm{R_{\odot}}$ &       11.6\\
\object{Kepler-42 b} &  $0.76 \mathrm{R_{\oplus}}$ & $0.17 \mathrm{R_{\odot}}$ &       11.5\\
\object{Kepler-42 c} &  $0.71 \mathrm{R_{\oplus}}$ & $0.17 \mathrm{R_{\odot}}$ &       11.5\\
\hline
\end{tabular}
\label{obstargets}
\end{table}

\section{Instruments} \label{Instruments}
High-resolution spectroscopy of exoplanetary atmospheres put stringent requirements on the instruments used for observations. We look into current, upcoming, and planned instruments that will be able to provide observations with sufficient S/N and spectral resolution. Most of the current high-resolution NIR spectrographs suffer from short wavelength coverage, which do not have high enough resolution or a large enough photon-collecting area. The otherwise impressive echelle spectrograph CRIRES at VLT (R = 100~000, 8 m primary) has a very narrow wavelength coverage and does not include a cross-dispersed mode to increase this, which makes the instrument hard to use with the method presented here. Other spectrographs with sufficient spectral coverage usually have spectral resolution well under 10~000, making observations with this method mostly pointless. There are, however, a few current instruments just at the lower edge of the needed requirements and several upcoming instruments that will be able to deliver observations with sufficient quality within the coming years.

\begin{figure}
\centering
\includegraphics[width=\hsize]{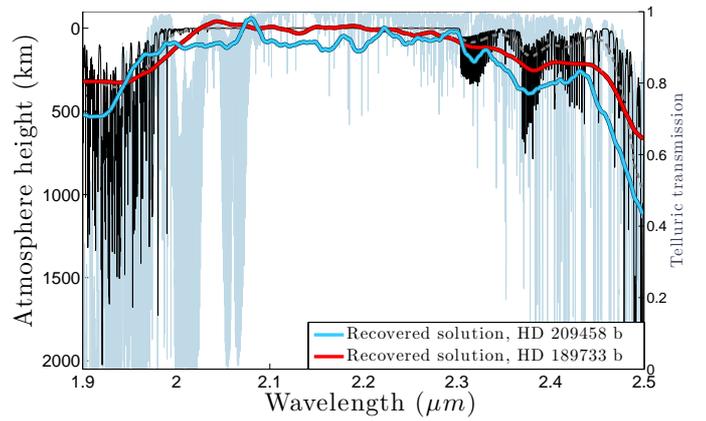} 
\caption{Recovered exoplanetary transmission spectra from simulated transit observations of systems with similar parameters to the two most promising observational candidates (see Table \ref{obstargets}), using NIR spectrograph IRCS at the Subaru telescope.\newline
Colour coding is the same as in Fig. \ref{G2M4M4}.}
\label{ircs}
\end{figure}

The two best-suited current spectrographs are NIRSPEC \citep{McLean1998} at Keck II and IRCS \citep{Tokunaga1998} at the Subaru telescope. NIRSPEC has a spectral resolution of 25~000 and a simultaneous wavelength coverage of about about 0.2 $\mu m$. IRCS has similar capabilities, a slightly lower spectral resolution of 20~000, but the simultaneous wavelength coverage is better, around 0.4 $\mu m$ at the relevant wavelengths. These instruments are at the lower edge of the instrumental requirements, and characterization of gas giants transiting the brightest stars should be possible with these instruments. We show this by simulating observations from one transit of the hot-Jupiters \object{HD 189733 b} and \object{HD 209458 b} (see Table \ref{obstargets}), using an instrument with spectral resolving power R = 20~000 at an 8 m telescope. Results shown in Fig. \ref{ircs}. 

While these current instruments should be able to characterize the large planets transiting the brightest stars, reaching super-Earths will not be possible unless large amounts of observing time is dedicated to this task. Driven largely by the need for precise radial velocity measurements of cool dwarf stars to detect Earth-size exoplanets, a number of upcoming high-resolution NIR spectrographs are currently planned. Many of these will be located at 4 m class telescopes, which will have a hard time reaching the needed S/N for transit spectroscopy. A few of the proposed or in-progress instruments will, however, be located at bigger telescopes.

One such instrument is the HZPF \citep{Mahadevan2010}, a proposed NIR high-resolution spectrograph, which will have a spectral resolution of 80~000, a wavelength coverage of 0.9 - 1.7 $\mu m$ in a single exposure, and the photon collecting power of the Hobby-Eberly Telescope's (HET) primary 9 m mirror. At first glance, this instrument seems very promising, however, HET cannot be pointed towards any part of the sky, and one must instead wait for a star with a transiting planet to pass over the telescope at exactly the right time. This keeps HET from being well-suited to observing rare and time-limited phenomena like exoplanetary transits. Combing observations from several transits would also be very hard, which makes recovering exoplanetary contributions from all but the largest planets impossible.

The Subaru telescope already has a NIR spectrograph capable of characterizing the most favourable cases (IRCS). There are, however, plans to add new high-resolution options to this telescope. IRCS-HRU \citep{Terada2008} will increase the maximum possible resolution of IRCS up to 70~000 in wavelengths longer than 1.4 $\mu m$. Another instrument, specialized in NIR radial velocity measurements is the IRD \citep{Tamura2012}, which will have a resolution of 70~000 and a simultaneous spectral coverage of 1.2 - 1.85 $\mu m$. Finally there is WINERED \citep{Yasui2008} with a resolution of 100~000 and simultaneous spectral coverage 0.9 - 1.35 $\mu m$. All of these instrument have the needed requirements for exoplanetary transit spectroscopy with the method presented here, and they might even be able to reach some of the more favourable super-Earths if enough observing time has been granted.

\begin{figure}
\centering
\includegraphics[width=\hsize]{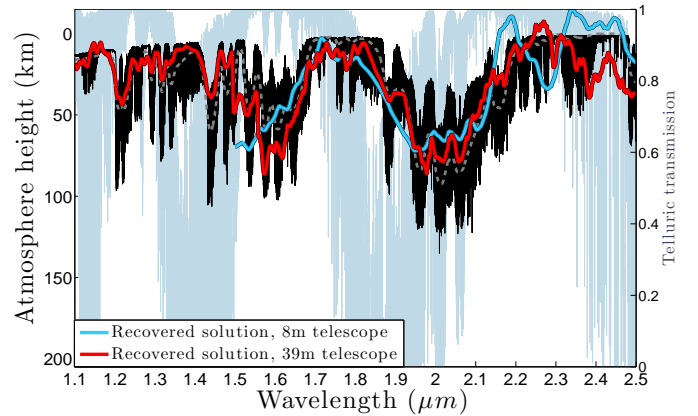} 
\caption{Recovered exoplanetary transmission spectra from simulated transit observations of a system with similar parameters GJ 1214 b (see Table \ref{obstargets}). Comparison between a 39m and 8m telescope.\newline
Colour coding is the same as in Fig. \ref{G2M4M4}.}
\label{HIRES}
\end{figure}

The current high-resolution NIR spectrograph at VLT, CRIRES, suffers from a very limited spectral coverage in a single exposure (~0.05 $\mu m$). There is a proposed upgrade for this instrument, CRIRES+ \citep{Oliva2012}. This instrument will have the same spectral resolution (100~000) and a significantly increased spectral coverage owing to the addition of a cross-dispersed mode. With the large photon collecting power of VLT's 8 m primary mirror and covering most of the Y, J, H, K, or M bands in a single exposure, observations from this instrument will be able to characterize many giant planets transiting bright stars. One should also be able to reach super-Earths transiting the very small stars if combing several transits. Advantageous super-Earths for these types of observations are, however, hard to find within reach of Paranal observatory (mainly due to all Kepler objects being located too far north), and currently the only known super-Earth where CRIRES+ observations will be sufficient for reproducing the transmission spectrum reliably is GJ 1214 b.

The NIR spectrographs for the next generation of 30 m class telescopes are even more impressive. For the 39 m European Extremely Large Telescope (E-ELT), the current idea is HIRES; a combined optical and NIR spectrograph with resolving power of at least 100~000 and simultaneous wavelength coverage of 1.0 - 2.7 $\mu m$. With the large photon collecting area of a 39 m mirror, observations during a single transits corresponds to observing more than 20 successive transits with an 8 m telescope. This makes characterizations of exoplanets orbiting faint stars feasible, as well as recovering high-resolution spectra from large planets around bright stars. (Results could be similar to Fig. \ref{Zoom in}.) The transit spectroscopy of Earth-size planets orbiting solar-like stars would still be hard because of the extremely small fraction of the stellar disk that is covered by the planetary atmosphere. To show the power of the photon collecting area of E-ELT, combined with a high-resolution spectrograph such as HIRES, we simulated observations of a single transit of \object{GJ 1214 b} (see Table \ref{obstargets}, assuming a atmospheric height of 150 km). This small planet orbits a faint star and would require observations of several transits with smaller telescopes to reach sufficient S/N, but with a 39 m mirror one would able to detect the main absorption bands from observations during a single transit. Results from this test are shown in Fig. \ref{HIRES}, including comparison with observations with a 8~m telescope. 

The Thirty Meter Telescope (TMT) also has planned a high-resolution NIR spectrograph, NIRES. This echelle spectrograph will be able to achieve spectral resolution up to 100~000 and will operate in 1 - 5 $\mu m$. A cross-dispersed mode will enable observations from 1.0 - 2.5 $\mu m$ simultaneously; as a result, apart from a smaller photon collecting area, the performance of NIRES will be very similar to E-ELT's HIRES. However, since the TMT is planned to be built on Mauna Kea, the telescopes will have access to the northern sky, where currently the majority of the favourable super-Earths and Earth-size exoplanets for transit spectroscopy are located.

\section{Observations} \label{CRIRESobservations}
To test the reduction procedure on real data we applied the data analysis method to CRIRES observations of \object{HD 209458 b} \citep{Snellen2010}. Data consists of 51 spectra (30 during transit) of a single three-hour transit of \object{HD 209458 b} (see Table \ref{obstargets}), targeting a CO band around 2.30 to 2.35 $\mu m$. We only used data from Detectors 2 and 3 owing to poor data quality in Detectors 1 and 4. In this region, there are few stellar lines, which reduces the need for accurate specific intensity spectra. Instead we used quadratic limb darkening law to model changes in stellar intensity as function of limb distance. Flux spectra are taken from the out-of-transit observations. Model telluric spectra were created using similar procedure to those described by \citet{Snellen2010}, except that we do not assign zero weight to the pixels in which we expect exoplanetary signal; instead, we increase the weight of pixels during out-of-transit exposures.

\begin{figure}
\centering
\includegraphics[width=\hsize]{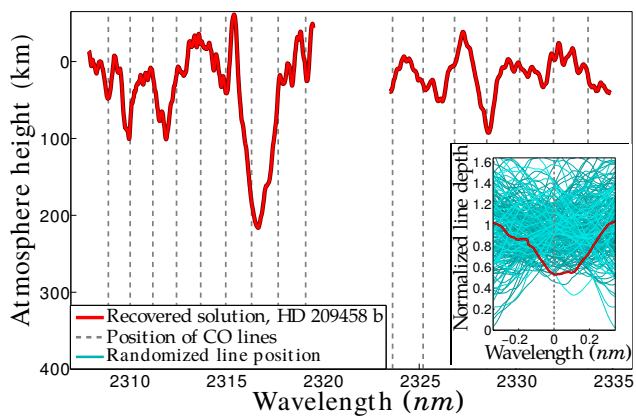}
\caption{Recovered exoplanetary transmission spectrum of \object{HD 209458 b} from CRIRES observations of a single transit. The expected positions of CO lines are shown as dashed grey lines. Only a few of the 16 CO lines in the observed wavelength range are visible in the recovered spectrum.\newline
Inserted panel: Stacked exoplanet spectra (normalized to average value in wings) over the expected position of CO lines (shown in red) and small sample (200) of 100~000 realizations of randomized positions of line centres (shown in turquoise). Only 1\% of the randomized line positions show stronger line depth than the CO lines.}
\label{resultsCRIRES}
\end{figure}

Owing to the short wavelength range, we cannot use the optimal regularization parameter based on data quality and expected signal strength, because doing so would result in a flat featureless solution. Instead we decrease regularization to what would be appropriate for recovering individual absorption lines in spite of the data not having the needed quality to recover individual lines. The recovered solution is shown in Fig. \ref{resultsCRIRES}, along with the placement of the CO lines previously detected in the same data set. The solution appears noisy, and only a few individual CO lines seem to be reconstructed. From this data it is not possible to directly identify the absorbers in the atmosphere of \object{HD 209458 b}. However, analysis of multiple CO lines will increase the signal. Stacking the available CO lines (ignoring the lines at 2317.7 and 2323.6 $nm$ due to several bad pixels close to line cores) and normalizing by the average value in the wings reveals an overall absorption feature centred on the core of the combined CO lines (see inserted panel in Fig. \ref{resultsCRIRES}). To show that the recovered feature is not a coincidence, we also stack and normalize the recovered solution over randomized line positions. From 100~000 different realizations of such randomized line positions, only 1\% shows absorption at the core of the line stronger than what we get from combing positions of CO lines.

\section{Conclusions} \label{Conclusions}
The data analysis method presented here makes the recovery of exoplanetary atmospheric transmittance spectrum using high-resolution observations of transiting exoplanets robust. There are several advantages to this method over the classical narrow band filter photometry approach. One of the most important aspects is the interpretation of the data, which becomes more straightforward thanks to higher spectral resolution and a wide wavelength coverage. Given the estimates of the physical conditions in the planetary atmosphere (pressure, temperature regime, UV-flux, etc.), a first approximation of chemical composition could be made directly from the recovered spectrum, without much need for atmospheric models. 

Molecular species responsible for strong absorption features (such as H$_2$0, CO$_2$, CO, CH$_4$) could be identified from the shape, placement, line spacing, and relative strength of recovered absorption bands. With broad enough spectral coverage to detect multiple absorption bands from a single molecular species, false detections could be minimized, and they would enable us to distinguish between species with overlapping absorption bands. From such a first approximation of the chemical composition, the next step would be to create a grid of planetary models with a range of relative abundances and temperature and pressure structures. Finding the model with the best fit to observations could now be made with fewer uncertainties, mainly due to the first approximation of chemical composition already made, but also due to the large number of data points that come with good spectral resolution and wide wavelength coverage. 

This method requires a special set of observations. To optimize observations, many short exposures (around 1-2 minutes) of the star are needed while the planet transits. Along with this, accurate stellar flux spectrum (without transiting planet), specific intensities and telluric transmittance, corresponding to each individual exposure, are all needed. In order not to reduce the observing time of the actual target, which is limited by the transit duration, we propose using synthetic telluric spectra fitted to observations. We estimate that the requirements of instruments and observational targets are able to produce observations with sufficient quality. Instruments need high spectral resolution (>20~000), good simultaneous spectral coverage (>0.1 $\mu m$), and a large photon collecting area (>8 m). 

We have identified potentially problematic issues. Large errors in telluric transmission spectra and stellar specific intensity can present as artificial features in the recovered exoplanet spectrum. We propose ways to minimize these problems and show that even under non-optimal conditions recovering the exoplanet spectrum is still possible.

For target selection, we defined the important aspects of targets and produced a short list of the most advantageous star-planet systems known to date (November 2014). The critical property of a target when performing transit spectroscopy is the relative size of the planet to its star. This makes the atmospheres of large gas giants generally much easier to recover. For spectral characterization of Jupiter-size exoplanets, this data analysis method using observations with high-resolution spectrometers at 8 m class telescopes seem promising. Recovering transmission spectra with this method will be possible once enough instrumentshave been constructed for most known gas giants orbiting bright stars. The chemical composition of these worlds might give us an insight into their origins. 

Smaller planets, super-Earths, and Earth-size planets are more difficult to characterize. We expect to be able to distinguish spectral features originating in their atmospheres given sufficient observing time, accurate stellar models, and advantageous properties of the host star (small and bright). Unfortunately, very few favourable candidates are known, and most are located out of the reach of two of the most promising upcoming instruments, CRIRES+ at VLT and HIRES at E-ELT. However, the success of the Kepler space missions has demonstrated that the chances of finding observationally advantageous super-Earths on the southern hemisphere is high, so all we need are dedicated space-based transit surv	eys. 

\begin{acknowledgements}
This research made use of the Exoplanet Orbit Database and the Exoplanet Data Explorer at exoplanets.org.\\
NSO/Kitt Peak FTS data used here were produced by NSF/NOAO.
\end{acknowledgements}

\bibliography{using_near_infrared.bib}
\bibliographystyle{aa}

\end{document}